\documentclass[prb,twocolumn,showpacs,preprintnumbers]{revtex4}

\usepackage{amsmath}
\usepackage{amssymb}
\usepackage{graphicx}

\newcommand{\gC}{{\check G_c}}
\newcommand{\cI}{{\check I}}
\newcommand{\cG}{{\check G}}
\newcommand{\one}{{1}}
\newcommand{\cp}{{\check p}}
\newcommand{\cc}{{\check c}}

\newcommand{\cga}{{\check \alpha}}
\newcommand{\cgb}{{\check \beta}}
\renewcommand{\c}{\check}

\newcommand{\calG}{{\check {\cal G}}}
\newcommand{\cE}{{\cal E}}
\newcommand{\cT}{{\cal T}}

\newcommand{\ctK}{{\check \tau_K}}

\newcommand{\Tn}[1]{{T_n^{(#1)}}}
\newcommand{\com}[2]{{[#1,#2]}}
\newcommand{\aCom}[2]{{\{#1,#2\}}}

\newcommand{\Tr}{{\mathop{\rm Tr}\nolimits}}

\newcommand{\fNp}{{f_N^+}}
\newcommand{\fNm}{{f_N^-}}
\newcommand{\fNpT}{{\tilde f_N^+}}
\newcommand{\fNmT}{{\tilde f_N^-}}

\begin{document}


\title{Full counting statistics of Andreev scattering in an asymmetric
  chaotic cavity}

\author{Mihajlo Vanevi\' c}
\author{Wolfgang Belzig}%
\affiliation{Departement f\" ur Physik und Astronomie,
Klingelbergstrasse 82, 4056 Basel, Switzerland}

\date{\today}

\begin{abstract}
  We study the charge transport statistics in coherent two-terminal
  double junctions within the framework of the circuit theory of
  mesoscopic transport. We obtain the general solution of the
  circuit-theory matrix equations for the Green's function of a chaotic
  cavity between arbitrary contacts. As an example we discuss the full
  counting statistics and the first three cumulants for an open
  asymmetric cavity between a superconductor and a normal-metal lead at
  temperatures and voltages below the superconducting gap. The third
  cumulant shows a characteristic sign change as a function of the
  asymmetry of the two quantum point contacts, which is related to the
  properties of the Andreev reflection eigenvalue distribution.
\end{abstract}




\maketitle

\section{Introduction}

Mesoscopic heterojunctions with superconductor (S) and
normal-metal or semiconductor (N) leads exhibit a rich variety of
phenomena that have been studied through conductance and noise for
some time already\cite{art:BTK,art:VolkovZaitsev,art:Beenakker92,%
art:Khlus,art:deJongBeenakkerPRB94,art:ThMartin,art:Datta,%
art:Blatter98,art:Imry01,art:Lambert,art:Takane,%
art:BuettikerPRB01,art:PistolesiPRB01}; see also
Refs.~\onlinecite{art:BuettikerPRB92,art:deJongBeenakker,%
  art:BeenakkerRevModPhys,art:BlanterButtikerPhysRep00,nazarov:book}.
In the absence of disorder charge transport through an interface
between S and N can be described by the Blonder-Tinkham-Klapwijk
(BTK) model,\cite{art:BTK} while the effects of scattering by
impurities and interface roughness can be taken into account using
the scattering theory
of mesoscopic transport.\cite{art:Beenakker92,art:deJongBeenakkerPRB94,%
art:ThMartin,art:Lambert,art:BuettikerPRB92,art:deJongBeenakker,%
art:BeenakkerRevModPhys,art:BlanterButtikerPhysRep00}
Recent experimental advancements\cite{art:SchoelkopfPRL97,%
art:OberholzerPRB99,art:SchoenenbergerPRL01,art:JehlNature00,%
art:ProberPRL00,art:HoffmannPRL03,art:Choi,art:ProberPRL03,art:Bomze}
raised interest in the statistical properties of the charge
transfer in mesoscopic systems. Current noise contains information
on the temporal correlation of quasiparticles relevant for
transport and originates from the fluctuations of quasiparticle
occupation numbers or random scattering at barriers or impurities.
Further, it can be used to probe internal energy scales of the
system or the effective charge of the elementary transport
mechanisms,\cite{art:deJongBeenakker,art:BeenakkerRevModPhys,%
art:BlanterButtikerPhysRep00,nazarov:book}
as well as to detect the correlations intrinsic
to the many-body state of entangled systems.\cite{nazarov:book}

If the charge transfer events are uncorrelated, the zero-frequency
noise power at low temperatures (shot noise) has the Poisson value
$P_I=2eI$, where $e$ is electron charge and $I$ is the average
current through the sample. This is the case for the charge
transport in a low-transparency normal-state tunnel junction at
zero temperature and low bias.  In less-opaque normal-state
junctions shot noise is usually suppressed below the Poisson
value, with the deviations characterized by the Fano factor
$F=P_I/2eI=[\sum_n T_n(1-T_n)]/\sum_n T_n$, where $T_n$ are
transmission eigenvalues of the junction.  For a metallic
diffusive wire\cite{art:SchoelkopfPRL97,art:OberholzerPRB99} and
for an open symmetric chaotic cavity\cite{art:SchoenenbergerPRL01}
Fano factors have universal values\cite{art:BeenakkerRevModPhys,%
  art:NazarovPRL94,art:SukhorukovPRL98,art:BlanterPRL00}
$F=1/3$ and $F=1/4$, respectively, which do not depend on
microscopic properties like the impurity concentration or geometry
of the sample.

The effect of superconductivity leads to a doubled shot noise in
subgap transport through low-transparency S/N tunnel
junctions\cite{art:HoffmannPRL03} and in diffusive normal wires in
contact with a
superconductor.\cite{art:JehlNature00,art:ProberPRL00} This
doubling can be understood as a consequence of the effective
charge doubling in the Andreev process.\cite{art:Andreev} In
general, however, the noise is also affected by the change of
transmission properties of the structure due to the superconductor
proximity effect and the doubling is not generic.  For example, in
an open symmetric cavity with a superconducting and normal-metal
lead the Fano factor has the value\cite{art:deJongBeenakker}
$F\approx 0.60$ which is more than two times larger than in the
corresponding normal-state junction ($F=1/4$), in agreement with
recent experimental results.\cite{art:Choi} At bias voltages on
the order of the superconducting gap both normal and Andreev
scattering processes contribute to
transport\cite{art:BelzigInQNoise,art:Muzykantskii} and the
picture of the effective charge carriers
fails.\cite{art:LesovikPRB01}

The statistical theory of mesoscopic
transport,\cite{art:LevitovJETP} full counting statistics,
provides the complete characterization of the charge transfer and
has led to a new and fundamental understanding of quantum
transport phenomena in nanoscale
conductors.\cite{art:LevitovYakovets,art:Muzykantskii}
Higher-order moments of the charge transport statistics may
provide additional information to the current noise
measurements.\cite{art:LevitovReznikovCondMat01,art:Bulashenko,art:LevitovJETP}
The third-order correlations of voltage fluctuations across the
nonsuperconducting tunnel junctions have been measured by Reulet
\textit{et al.}\cite{art:ProberPRL03} and recently by Bomze
\textit{et
  al.},\cite{art:Bomze} the latter confirming the Poisson statistics of
the charge transfer at negligible coupling of the system to environment.
The semiclassical theory of higher-order cumulants based on the
Boltzmann-Langevin equations has been developed recently by Nagaev
\textit{et al.}\cite{art:Nagaev}
The quantum-mechanical approach to full counting statistics based
on the extended Keldysh-Green's function
technique\cite{art:nazarovANNPHYS99,art:BelzigPRL01} in the
discretized form of the circuit
theory\cite{art:Nazarov95,art:NazarovSuperlatt99} was put forward
for multiterminal circuits by Nazarov and
Bagrets.\cite{art:NazarovBagretsPRL02} This approach can describe
junctions with different connectors and leads as well as
multiterminal circuits in unified and very general way.  Within
the circuit theory, the doubling of shot noise (i.e., $F=2/3$) was
found in diffusive S/N junctions in both the fully coherent $eV\ll
E_{Th}$ and the completely incoherent $eV\gg E_{Th}$
regimes,\cite{art:BelzigSamuelssonEPL03,belzig:01-1} where
$E_{Th}$ is the inverse diffusion time.  At intermediate bias
voltages $eV\sim E_{Th}$, the transport is affected by
electron-hole coherence leading to an enhancement of the
differential Fano factor.\cite{belzig:01-1,art:belzigPRL04} The
influence of coherence effects on the full counting statistics and
noise in other single and double S/N junctions was studied in
Refs.~\onlinecite{art:SamuelssonPRB03,art:HouzetPistolesiPRL04,art:Pistolesi}.
The current correlations in a three-terminal chaotic cavity
operated as superconducting beam splitter were studied by
Samuelsson and B\" uttiker,\cite{art:SamuelssonButtikerPRL02}
yielding unusual positive cross correlations. From a study of the
full counting statistics these positive correlations can be
attributed to the uncorrelated injection of Cooper
pairs.\cite{boerlin:02}

In this article we study the full counting statistics of coherent
charge transport in a chaotic cavity using the circuit theory of
mesoscopic transport. We show that the system of matrix equations
for the Green's function of the cavity can be solved, effectively,
as a system of scalar equations independently of the type of leads
and without resorting to the matrix components or
parametrizations.  As an application we find the Green's function
for an open asymmetric cavity between arbitrary leads.  For the
special case of a cavity between the superconductor and normal
metal, we find the cumulant generating function and the first
three cumulants and discuss the interplay between superconducting
proximity effect and scattering properties of the junction. The
results are compared with those for a normal-state
junction\cite{art:Bulashenko} and for different couplings of a
cavity to the leads. Current correlations in a structure with
high-quality contacts between a cavity and superconductor have
been studied experimentally by Choi {\it et al.}\cite{art:Choi}
recently.

\section{Model}

\begin{figure}[b]
\includegraphics[width=5.4cm, height=4cm]{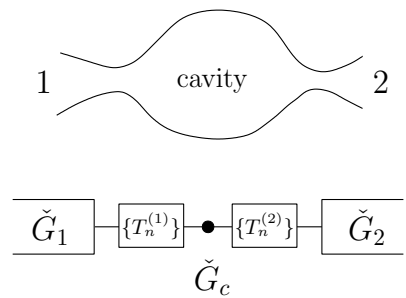}
\caption{\label{fig:Scheme} A chaotic cavity coupled to the leads
  by two junctions with transmission eigenvalues $\{\Tn{1}\}$ and
  $\{\Tn{2}\}$ (top). The discrete circuit-theory representation of the
  system is shown in the lower part. The leads and cavity are
  characterized by the corresponding matrix Green's functions. The
  junctions, depicted as connectors, carry conserved matrix currents.}
\end{figure}

The system we consider is a chaotic cavity coupled to two leads by
mesoscopic junctions characterized by the transmission
eigenvalues $\{T_n^{(1)}\}$ and $\{T_n^{(2)}\}$, respectively.
Charging of the cavity is negligible if the cavity is
large enough and the conductances of
the junctions $g_{1,2}$ are much larger than the conductance
quantum $g_0=2e^2/h$. We assume an isotropic quasiparticle
distribution function inside the cavity due to chaotic scattering.
The decoherence effects as well as the energy dependence of
transmission eigenvalues can be neglected if the total dwell time
in the cavity is small with respect to time scales set by the
inverse temperature and bias voltage. We apply the circuit theory
of mesoscopic transport\cite{art:NazarovSuperlatt99} with the
specific parts of the system represented by the corresponding
discrete circuit elements, as shown in Fig.~\ref{fig:Scheme}.
The leads are characterized by known quasiclassical matrix Green's
functions $\cG_{1,2}$ which depend on the quasiparticle energy,
lead temperature, chemical potential, and counting field and
satisfy the normalization condition $\cG_1^2=\cG_2^2=\one$. The
lead Green's functions are not necessarily the ones of bulk
electrodes --  for example, they can be nodes that are part of a
larger circuit. In the following we will refer to these nodes as
leads. The formulation below is independent of the concrete matrix
structure, provided the 'lead' Green's functions obey the
normalization conditions. The chaotic cavity is represented as an
internal node associated with an unknown Green's function $\gC$,
which will be obtained from the matrix current conservation and
the normalization condition $\gC^2=\one$. Left and right junctions
($i=1,2$), depicted as connectors, carry matrix
currents\cite{art:NazarovSuperlatt99}
\begin{equation}\label{eq:matrixCurr}
\cI_i = \frac{2e^2}{h}
\sum_n
\frac{2\Tn{i}\com{\cG_i}{\cG_c}}{4+\Tn{i}(\aCom{\cG_i}{\cG_c}-2)},
\end{equation}
which flow from the cavity into the leads. The current
conservation $\cI_1+\cI_2=0$ for the Green's function $\gC$ of the
cavity reduces to
\begin{equation}\label{eq:eqGc}
\left[
\cp_1\cG_1 +
\cp_2\cG_2,
\gC
\right]=0,
\end{equation}
with
\begin{equation}\label{eq:p12}
\cp_i = \sum_n \frac{\Tn{i}}{4+\Tn{i}(\aCom{\cG_i}{\cG_c}-2)}.
\end{equation}
Here we have used the commutation of $\cp_{1(2)}$ with
$\cG_{1(2)}$ and $\gC$, which is a consequence of the
normalization $\cG_i^2=\cG_c^2=1$, and the matrix property
\begin{equation}\label{eq:matrixProp}
\check A^2=\one \quad \Rightarrow \quad
\com{\check A}{\aCom{\check A}{\check B}}=0.
\end{equation}

We can solve Eq.~\eqref{eq:eqGc} while assuming that $\cp_1$ and
$\cp_2$ depend only on the anticommutator of the Green's functions
of the leads, $\cp_i=\cp_i(\aCom{\cG_1}{\cG_2})$ (see the
Appendix), and commute with $\cG_1$, $\cG_2$, and $\cG_c$ in
accordance with Eq.~\eqref{eq:matrixProp}. As a result, the
Green's function of the cavity can be expressed in terms of
Green's functions of the leads in the form
\begin{equation}\label{eq:GcRes}
\cG_c = \frac{\cp_1}{\cc} \cG_1 + \frac{\cp_2}{\cc} \cG_2,
\end{equation}
where the matrix $\cc=\cc(\aCom{\cG_1}{\cG_2})$ accounts for the
normalization of $\cG_c$. From Eq.~\eqref{eq:GcRes} and by using
the normalization conditions $\gC^2=\cG_1^2=\cG_2^2=\one$ we
obtain the system of equations
\begin{gather}\label{eq:system1}
\cc^2 = \cp_1^2 + \cp_2^2 + \cp_1\cp_2 \calG\,, \\
\cc\, \calG_1 = 2\cp_1 + \cp_2 \calG\,, \\
\cc\, \calG_2 = 2\cp_2 + \cp_1 \calG\,,\label{eq:system3}
\end{gather}
where $\calG_i=\aCom{\cG_i}{\gC}$ and $\calG=\aCom{\cG_1}{\cG_2}$.
This system can be treated, effectively, as a system of scalar equations
because all matrices that appear in
Eqs.~\eqref{eq:system1}--\eqref{eq:system3} depend only on $\calG$
and commute with each other.

The cumulant generating function $S(\chi)$ of charge transfer can
be obtained as a sum of the actions of the connected pairs of
nodes\cite{art:NazarovBagretsPRL02}:
\begin{equation}\label{eq:Sgen}
S(\chi) = -\frac{t_0}{2h} \sum_{i=1,2} \sum_n \int d\cE
\; \Tr\ln\Big(1+\frac{\Tn{i}}{4}(\calG_i-2)\Big),
\end{equation}
where the total measurement time $t_0$ is much larger than the
characteristic time scale on which the current fluctuations are
correlated. The $\chi$-independent term in the cumulant generating
function [given by the normalization requirement $S(\chi=0)=0$] is
omitted for brevity throughout this article. Also, we recall that
because of current conservation, it is sufficient to introduce a
counting field $\chi$ at one lead only, while the full counting
field dependence can be obtained by setting $\chi=\chi_1-\chi_2$.

The cumulant generating function $S$ depends only on the
anticommutator $\calG(\chi)$ and is invariant to the exchange of
the leads $\cG_1 \leftrightarrow \cG_2$ or, equivalently, to
exchange of the junctions $\{\Tn1\} \leftrightarrow \{\Tn2\}$ [see
Eqs.~\eqref{eq:eqGc} and \eqref{eq:p12}].  Therefore, the same
invariance
persists\cite{art:BelzigInQNoise,art:Bulashenko,art:Pistolesi} in
all coherent (or low bias) transport properties of two-terminal
double junctions -- such as current (conductance), noise (Fano
factor), and higher cumulants -- independently of the type of
leads or specific properties of the junctions. This invariance
does not hold in the presence of
dephasing,\cite{art:SamuelssonPRB03,art:Pistolesi} which can be
modeled by an additional lead that carries the coherence leakage
current.\cite{art:NazarovSuperlatt99,art:deJongBeenakker,art:BeenakkerRevModPhys,art:BlanterButtikerPhysRep00}

In the following we consider an analytically tractable case of a
chaotic cavity coupled to the leads by two quantum point contacts
with $N_1$ and $N_2$ open channels, respectively. The transmission
eigenvalues of the contacts are $\Tn{i}=1$ for $n=1,\ldots,N_i$
and $\Tn{i}=0$ otherwise. From Eq.~\eqref{eq:p12} and
Eqs.~\eqref{eq:system1}--\eqref{eq:system3} we find
\begin{equation}\label{eq:GcRes2}
\cc
=
\frac{N_1+N_2}{2} \left( 1+\sqrt{1-\frac{4
N_1N_2}{(N_1+N_2)^2}\frac{\calG-2}{\calG+2}} \right)^{-1}
\end{equation}
and
\begin{equation}\label{eq:GcRes3}
\frac{\cp_{1,2}}{\cc}
=
\frac{1}{2}
\left(
\pm \frac{N_1-N_2}{N_1+N_2} +
\sqrt{1-\frac{4 N_1N_2}{(N_1+N_2)^2}\frac{\calG-2}{\calG+2}}
\right).
\end{equation}
The Green's function of the cavity and cumulant generating
function are given by Eqs.~\eqref{eq:GcRes} and \eqref{eq:Sgen},
respectively.  A formally similar result has been obtained
recently by Bulashenko\cite{art:Bulashenko} using $2\times 2$
Green's functions in Keldysh space which can be expanded over the
Pauli matrices. Physically, this implies that the whole circuit is
in the normal state, although it is permitted that the leads are
nodes of a larger (normal-state) mesoscopic network. In our
approach we do not rely on this expansion and make no assumptions
on the particular matrix structure, except for the usual
normalization condition.  Therefore, Eqs.~\eqref{eq:GcRes2} and
\eqref{eq:GcRes3} are valid for {\em any type of leads}.  For
example, one lead can be superconducting, with the Green's
function having Keldysh-Nambu matrix structure, or the chaotic
cavity can be a part of the larger multiterminal network which
consists of different heterojunctions. Additional degrees of
freedom -- for instance, spin -- can be included as well. We
emphasize, again, that our solution only resorts to the
normalization condition of the leads. In the case in which the
cavity is part of a larger network, Green's functions $\cG_{1,2}$
have to be determined by circuit rules. The result for $\cG_c$ is
valid in this case as well, which can simplify the numerical
solution of larger circuits.

It is interesting to check that an alternative approach can give the
same result. Coherent connectors in the circuit theory are described
by a cumulant generating function of the form
\begin{equation}\label{eq:SgenLevitov}
S(\chi)= -\frac{t_0}{2h}\int d\cE \int_0^1 d\cT \rho(\cT)
\; \Tr\ln \Big( 1+\frac{\cT}{4}[\calG(\chi)-2]\Big),
\end{equation}
where $\rho(\cT)$ is the distribution of
transmission eigenvalues $\{T_n\}$ for the composite junction.
Using the distribution of transmission eigenvalues for an open
chaotic cavity,\cite{art:Nazarov95,art:BeenakkerRevModPhys}
\begin{equation}\label{eq:rhoTqpc}
\rho_c(\cT) = \frac{\sqrt{N_1N_2}}{\pi}\; \frac{1}{\cT}
\sqrt{\frac{\cT-\cT_0}{(1-\cT)(1-\cT_0)}},
\end{equation}
for $\cT_0<\cT<1$ and $\rho_c(\cT)=0$ otherwise, with
$\cT_0=(N_1-N_2)^2/(N_1+N_2)^2$, we obtain $S(\chi)$ as given by
Eqs.~\eqref{eq:Sgen}--\eqref{eq:GcRes3}. This demonstrates the
consistency of the circuit-theory approach with the random matrix
theory of scattering matrices.

\section{Superconductor -- chaotic-cavity -- normal-metal junction}

In the following we calculate the statistics of charge transport
through a chaotic cavity sandwiched between a superconductor and a
normal-metal lead. We present a detailed analysis of the first
three cumulants -- current, current noise power, and the third
cumulant of the current -- for an open chaotic cavity at
temperatures and bias voltages well below the superconducting gap
$\Delta$, when Andreev scattering is the dominant process of the
charge transfer.
At low energies and temperatures the $4\times 4$ matrix
Green's function of the superconductor is
$\cG_S\equiv\cG_1=\bar\one\;\hat\sigma_1$ in the
Keldysh($\bar{\;}$)$\otimes$Nambu($\hat{\;}$) space. The Green's
function $\cG_N=\cG_N(\cE,\chi)\equiv\cG_2$ of the normal-metal
lead incorporates the counting field according to
\begin{equation}
\cG_N = e^{-i\frac{\chi}{2}\ctK}\; \cG_N^0\;
e^{i\frac{\chi}{2}\ctK},
\end{equation}
where  $\ctK = \bar\tau_1 \hat\sigma_3$ and $\bar{\tau}_i$ and
$\hat{\sigma}_i$ are Pauli matrices. The bare Green's function of
the normal-metal lead is given by
\begin{equation}\label{eq:gN0}
\cG_N^0 =
    \begin{pmatrix}
    \hat\sigma_3 & 2\hat K \\
    0 & - \hat\sigma_3
    \end{pmatrix},
\quad
\hat K =
    \begin{pmatrix}
    1-2\fNp & 0 \\
    0 & 1-2\fNm
    \end{pmatrix},
\end{equation}
where $f_N^\pm = \{\exp[(\pm\cE + eV)/k_BT]+1\}^{-1}$ accounts for
the voltage bias of the normal-metal lead, with the energy $\cE$
measured in respect to the chemical potential of the
superconductor. From Eqs.~\eqref{eq:Sgen}--\eqref{eq:GcRes3}, we
find the following expression for the cumulant generating
function:
\begin{multline}\label{eq:CGFres}
S(\chi) = -\frac{t_0}{h} \int d \cE \sum_{j=1,2}
N_j
\\
\times \ln [ r_j + \sqrt{r_j^2-64 N_j^4(1+a)} ],
\end{multline}
where
\begin{multline}
r_{1(2)}=  a(N_1-N_2)^2+(3N_{1(2)}+N_{2(1)})^2
\\
+
\sqrt{(1+a)[a(N_1-N_2)^4+(N_1^2+N_2^2+6N_1N_2)^2]}\,.
\end{multline}
Here, $a={(e^{2i\chi}-1)\fNpT\fNmT + (e^{-2i\chi}-1)\fNp\fNm}$ and
$\tilde f_N^\pm = 1-f_N^\pm$, with $a$ being related to the
double-degenerate eigenvalues of $\calG=\aCom{\cG_S}{\cG_N}$ given
by $\lambda_{1,2}=\pm 2i\sqrt{\smash[b]{a}}$. From
Eq.~\eqref{eq:CGFres} we obtain the average current, the current
noise power, and the third cumulant according to $I=i (e/t_0)
\partial_\chi S|_{\chi=0}$, $P_I=(2e^2/t_0)\partial_\chi^2
S|_{\chi=0}$, and $C_I=-i(e^3/t_0)\partial_\chi^3S|_{\chi=0}$,
respectively. They are
\begin{align} \label{eq:Iav}
I =& \frac{G_S}{2e}\int d\cE \; (\fNpT\fNmT - \fNp\fNm),
\\ \label{eq:PI}
P_I =& 2 G_S \! \int d\cE \;
    [
    (\fNpT\fNmT+\fNp\fNm)-
    \gamma_1(\fNpT\fNmT-\fNp\fNm)^2 ],
\end{align}
and
\begin{multline} \label{eq:CI}
C_I = 2e G_S \int d\cE \; (\fNpT\fNmT-\fNp\fNm)
    \\
    \times
    [
    1-3\gamma_1(\fNpT\fNmT+\fNp\fNm)
    +2\gamma_2 (\fNpT\fNmT-\fNp\fNm)^2
    ],
\end{multline}
with the total conductance of S-cavity-N junction,
\begin{equation}
G_S=\frac{2e^2}{h}(N_1+N_2)\left(1-\frac{N_1+N_2}{q}\right),
\end{equation}
and the low-temperature Fano factor\cite{art:deJongBeenakker}
\begin{equation}
F_S=\frac{16N_1^2N_2^2(N_1+N_2)}{q^4[q-(N_1+N_2)]}\,.
\end{equation}
Here $q=\sqrt{\smash[b]{(N_1+N_2)^2+4N_1N_2}}$, $\gamma_1=1-F_S/2$,
and $\gamma_2=1-F_S[1-2N_1N_2(2N_1+N_2)(N_1+2N_2)/q^4]$.
After the energy integration in Eq.~\eqref{eq:Iav}, the usual
relation $I=G_SV$ is obtained, while the integration in
Eqs.~\eqref{eq:PI} and \eqref{eq:CI} yields
\begin{align} \label{eq:PIint}
\frac{P_I}{4G_Sk_BT} &= 1+(F_S/2)\;[v\coth(v)-1]
\\ \label{eq:PIintApprox}
&=  \begin{cases}
    1+(F_S/6)\; v^2, & |v|\ll 1, \\
    1-F_S/2 + (F_S/2)\; |v|, & |v|\gg 1,
    \end{cases}
\end{align}
and
\begin{align} \label{eq:CIint}
\frac{C_I}{eG_Sk_BT} =& 12(\gamma_1-\gamma_2)
    \bigg[
    \coth(v) \notag
    \\
    &\;\;\;\;\; + v\left(
    \frac{1-\gamma_2}{3(\gamma_1-\gamma_2)}-\coth^2(v)\right)
    \bigg]
\\ \label{eq:CIintApprox}
&\!\!\!\!\!\!\!\!\!\!\!\!\!\!\!\!
=   \begin{cases}
    2F_S \; v, & |v|\ll 1, \\
    \pm12(\gamma_1-\gamma_2)+4(1-3\gamma_1+2\gamma_2)\; v, & \pm v\gg 1,
    \end{cases}
\end{align}
respectively, with $v=eV/k_BT$.

The cumulant generating function, Eq.~\eqref{eq:CGFres}, takes
into account the superconductor proximity effect in the
quasiclassical approximation as well as interchannel mixing inside
the cavity in the presence of quantum point contacts. In
comparison with the normal-state junction (see
Ref.~\onlinecite{art:Bulashenko}),
Eqs.~\eqref{eq:Iav}--\eqref{eq:CI} contain products of electron
and hole distribution functions due to the Andreev process, which
is the mechanism of the charge transport. For example, the term
$\fNp\fNm=\fNp(1-\fNmT)$ can be interpreted as the probability for
an electron emerging from the lead N to be reflected back as a
hole, where $\fNp$ and $\fNmT$ are the electron- and hole-state
occupation numbers. The energy-independent prefactors $G_S$,
$\gamma_1$, and $\gamma_2$ are also modified by the electron-hole
correlations introduced by the superconductor. This change of
transport properties due to superconductor proximity effect can be
revealed by considering a general S/N junction with transmission
distribution $\rho(\cT)$, at low temperatures and bias voltages
($k_BT,|eV|\ll \Delta$). In this case Eq.~\eqref{eq:SgenLevitov}
reduces to
\begin{equation}\label{eq:Sqpc}
S(\chi)=-\frac{t_0}{2h}\int d\cE \int_0^1 dR_A \; \rho_A(R_A) \ln
[ 1 + R_A\; a(\chi) ],
\end{equation}
where $\lambda_{1,2}=\pm 2i\sqrt{a}$ are double-degenerate
eigenvalues of $\calG$. In the fully coherent regime, which we
consider here, the distribution of Andreev reflection eigenvalues
$\rho_A(R_A)$ is simply related to the distribution of
transmissions $\rho(\cT)$ of the corresponding normal-state
junction: $\rho_A(R_A)=2\rho(\cT) d\cT/dR_A$. The probability of
the Andreev reflection is given by $R_A=\cT^2/(2-\cT)^2$ and
appears in Eq.~\eqref{eq:Sqpc} due to electron-hole symmetry at
energies well below the superconducting gap and the inverse dwell
time in the cavity.
Normal scattering processes are suppressed, and $S(\chi)$ depends
on the counting field through $e^{\pm 2i\chi}$, which accounts for
the effective charge of $2e$ that is transferred across the
structure in each elementary event of Andreev scattering.  In the
case of strong electron-hole dephasing within the structure,
electrons and holes decouple and the system can be mapped onto an
effective normal-state
junction,\cite{art:BelzigSamuelssonEPL03,art:Pistolesi} for which
the cumulant generating function is given by
Eq.~\eqref{eq:SgenLevitov} with the corresponding modification of
transmission distribution $\rho(\cT)$ and boundary conditions
$\calG$. In the crossover regime, transport through the structure
is not simply related to the normal-state transmission properties
and can be described by an effective energy-dependent
distribution\cite{art:belzigPRL04} $\rho_A$ which takes into
account the effects of dephasing (coherence leakage currents) at
characteristic energies on the order of Thouless energy.

Expanding Eq.~\eqref{eq:Sqpc} in the $\chi$ field and taking the
derivatives, we obtain the current, current noise power, and the
third cumulant in the coherent regime as given by
Eqs.~\eqref{eq:Iav}--\eqref{eq:CI}, with conductance
\begin{equation}\label{eq:condRhoT}
\tilde G_S =2 \int_0^1 d\cT \rho(\cT) \; R_A,
\end{equation}
$\gamma_1 = 2\tilde G_S^{-1} \int_0^1 d\cT \rho(\cT) R_A^2$, and
$\gamma_2 = 2\tilde G_S^{-1} \int_0^1 d\cT \rho(\cT) R_A^3$, where
$\tilde G_S=G_S/(2e^2/h)$. In particular, the Fano factor and the slope
of the third cumulant at high bias are given by
\begin{equation}\label{eq:fanoRhoT}
F_S
=
\frac{4}{\tilde G_S} \int_0^1 d\cT \rho(\cT) \; R_A(1-R_A)
\end{equation}
and
\begin{equation}\label{eq:3cumRhoT}
\frac{\partial C_I}{\partial (e^2 I)}
=
\frac{8}{\tilde G_S} \int_0^1 d\cT \rho(\cT) \; R_A(1-R_A)(1-2R_A),
\end{equation}
respectively. These expressions are similar to the normal-state ones
except for the effective charge doubling and the presence of the Andreev
reflection probability $R_A$ instead of normal transmission $\cT$, in
agreement with previous results obtained within the scattering
approach.\cite{art:deJongBeenakkerPRB94} Using the transmission
distribution $\rho_c(\cT)$ given by Eq.~\eqref{eq:rhoTqpc}, we recover
the results for an open asymmetric cavity, which were obtained from the
circuit theory without the explicit knowledge of $\rho(\cT)$ for the
composite junction.

\begin{figure}[t]
\includegraphics{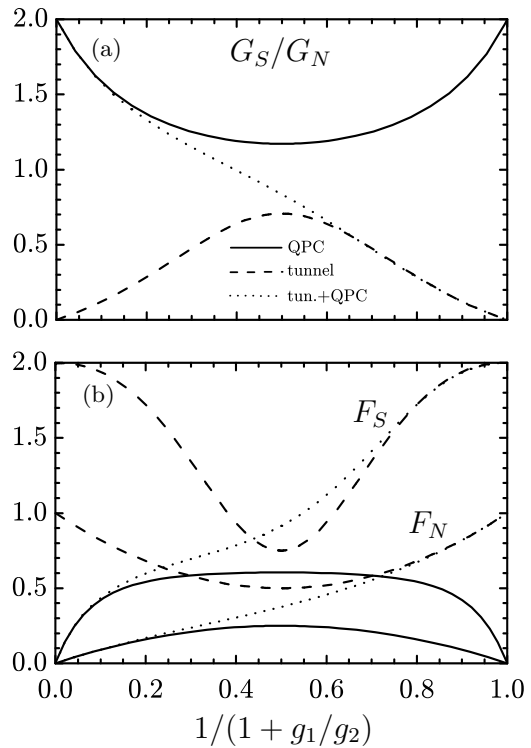}
\caption{\label{fig:ConductFano} Conductance $G_S$ [panel (a)] and
  the Fano factor $F_S$ [panel (b)] of the S-cavity-N junction as
  a function of the junction asymmetry $1/(1+g_1/g_2)$.
  Results for three different couplings of a cavity to the leads
  are shown for comparison: coupling by quantum point contacts
  (solid curves), tunnel junctions (dashed curves), and coupling by
  tunnel junction from the S side and quantum point contact from
  the N side (dotted curves). Conductance is
  normalized to the normal-state value $G_N=g_1g_2/(g_1+g_2)$,
  with $g_i=(2e^2/h)N_i$ for quantum point contacts
  and $g_i=(2e^2/h)\sum_n \Tn{i}$ for tunnel junctions.
  Fano factors $F_N$ for the corresponding normal-state junctions
  are shown in panel (b) for comparison.}
\end{figure}

The total conductance $G_S$ [normalized to the normal-state value
$G_N=g_1g_2/(g_1+g_2)$] and the Fano factor $F_S$ of the
S-cavity-N junction are shown in Fig.~\ref{fig:ConductFano} as
functions of the junction asymmetry and for different couplings
between the cavity and leads. For the symmetric
quantum-point-contact coupling $g_1/g_2=1$, the conductance ratio
has the minimal value $G_S/G_N=2(2-\sqrt{2})\approx 1.17$, while
the Fano factor is maximal, $F_S=(\sqrt{2}+1)/4\approx 0.60$. In
the highly asymmetric limit, $G_S/G_N = 2$ and
$F_S=8g_{min}/g_{max}\approx0$ (Fig.~\ref{fig:ConductFano}, solid
curves). The vanishing of the shot noise in this case is due to
the perfect transparency of the dominant (the one which is weakly
coupled) quantum point contact. For the case of two tunnel
junctions\cite{art:BelzigInQNoise,art:SamuelssonPRB03} instead of
quantum point contacts, the trend is opposite: for the symmetric
coupling, $G_S/G_N=1/\sqrt{2}\approx 0.71$ is maximal and
$F_S=3/4$ is minimal, and for the highly asymmetric coupling,
$G_S/G_N=g_{min}/g_{max}\approx0$ and $F_S=2$ (see
Fig.~\ref{fig:ConductFano}, dashed curves). We point out that
these different trends can be used to probe the quality of the
contacts.
The dotted curves in Fig.~\ref{fig:ConductFano} show numerical
results for the conductance and the Fano factor of a cavity
coupled to a superconductor by a tunnel junction and to a normal
lead by quantum point contact. We find that the transport
properties of the system are not affected by the type of coupling
to the superconductor when the quantum point contact on the normal
side dominates. This limit is reached at the conductance ratio
$g_1/g_2 \gtrsim 5$ of the tunnel- and quantum-point-contact
coupling of the cavity to the leads.  Therefore, the junction that
corresponds to the model of an open chaotic cavity can be realized
either with two good quality quantum point contacts from the both
sides or it can be an asymmetric junction with only the
normal-side quantum point contact of a high quality. The latter is
easier to fabricate, and the required asymmetry can be achieved by
increasing the contact area of the cavity to the superconductor.
Experimentally, the conductance and current noise power have been
measured recently by Choi {\it et al.},\cite{art:Choi} in a setup
which is very similar to the system we have analyzed. As the
estimates from Ref.~\onlinecite{art:Choi} show, it is possible to
fabricate a high-quality contact between cavity and
superconductor. The measured values of the Fano factors
$F_S=0.58\pm0.10$ and $F_N=0.25\pm0.04$ across the junction in the
superconducting- and normal-state regimes, respectively, are in
agreement with the model of symmetric open chaotic cavity [compare
with solid curves in Fig.~\ref{fig:ConductFano} (b)]. However, the
measured conductance ratio $G_S/G_N \approx 0.90$ is in
discrepancy with the conductance ratio $G_S/G_N=1.17$ predicted by
this simple model. The difference may originate from the inelastic
quasiparticle scattering at the disordered superconductor
interface, nonuniversal correction due to relatively large
openings of the cavity,\cite{art:BlanterPRL00} or dephasing of
quasiparticles due to an additional lead which is left floating in
the experiment.

At high temperatures $k_BT\gg |eV|$, current noise power is
thermally dominated and linear in conductance and temperature [see
Eq.~\eqref{eq:PIintApprox}], as expected from the
fluctuation-dissipation theorem. Thus, to extract the Fano factor
from the current noise power measurement, it is necessary to be in
the low-temperature, shot noise regime $k_BT\ll|eV|$.
Experimentally, this requires high bias voltages at which
nonlinear $I-V$ characteristics occur, especially in a strongly
interacting electron systems, with a difficulty how to distinguish
the shot noise contribution from the contribution of thermal noise
modified by nonlinear conductance.


\begin{figure}[t]
\includegraphics{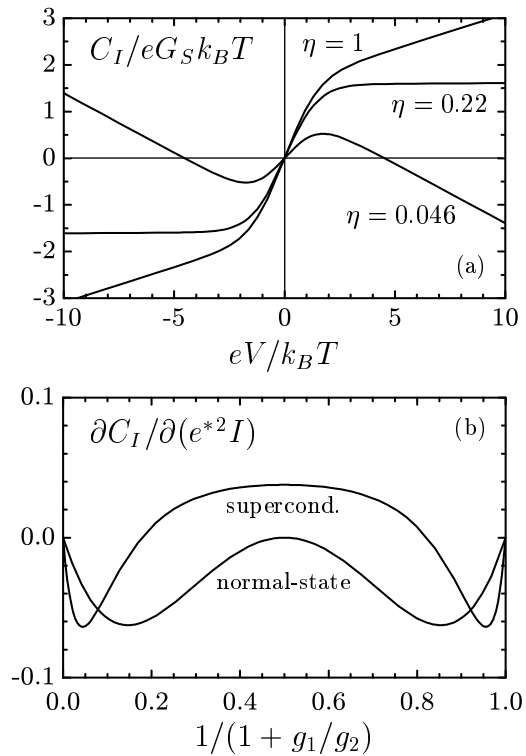}
\caption{\label{fig:ThirdCumulant}
  The third cumulant as a function of bias-to-temperature ratio
  [panel (a)], shown for three characteristic junction asymmetries:
  $C_I$ has the maximal positive slope at high biases for
  symmetric coupling $\eta=g_{min}/g_{max}=1$, saturates at high biases for
  $\eta\approx 0.22$, and has maximal negative slope for $\eta\approx
  0.046$. The high-bias slope of the third cumulant is shown
  in panel (b) as a function of the junction asymmetry, with
  the normal-state value given for comparison. Effective charge is $e^*=2e$
  and $e^*=e$ for the superconducting and normal-state junction, respectively.}
\end{figure}

\begin{figure}[t]
\includegraphics{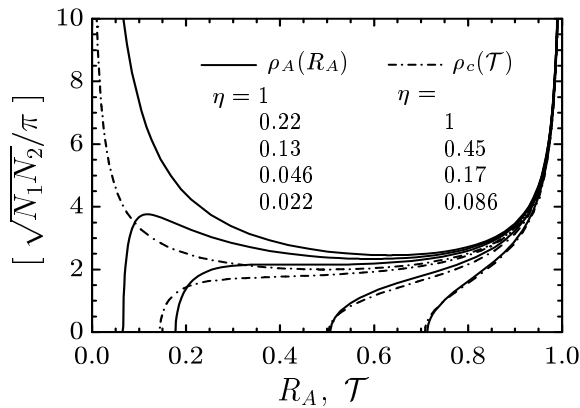}
\caption{\label{fig:Distribution}
  Distribution of the Andreev reflection probabilities
  $\rho_A(R_A)=2\rho_c(\cT) d\cT/dR_A$ (solid curves) and
  the distribution of transmission eigenvalues
  $\rho_c(\cT)$ (dash-dotted curves) for an open chaotic cavity,
  shown for different asymmetries of the junction
  $\eta=g_{min}/g_{max}$ (from top to bottom).}
\end{figure}

Finally, we point out the difference between the superconducting
and normal-state asymptotic behavior of the third cumulant at high
biases. For an asymmetric open cavity coupled to normal-metal
leads, the slope of the voltage-dependent third cumulant is
negative, reaching zero for the symmetric
cavity.\cite{art:Bulashenko} When one lead is superconducting,
this slope is negative for highly asymmetric couplings and
positive for symmetric couplings
[Fig.~\ref{fig:ThirdCumulant}~(b)], with the crossover at
$\eta=g_{min}/g_{max}=
\sqrt{\smash[b]{3+2\sqrt{2}}}-\sqrt{\smash[b]{2+2\sqrt{2}}}
\approx 0.22$. Thus, in the normal-state regime the third cumulant
changes sign at high enough biases, while in the superconducting
case this happens only if the junction is sufficiently asymmetric.
This difference originates from the interchannel mixing inside the
cavity in the presence of the superconducting proximity
effect\cite{art:BelzigInQNoise} and can be attributed to the
skewness\cite{art:Bulashenko} of the Andreev reflection
probability distribution function
$\rho_A(R_A)=2\rho_c(\cT)d\cT/dR_A$ [compare with
Eqs.~\eqref{eq:condRhoT}--\eqref{eq:3cumRhoT}]. For the
normal-state symmetric cavity, the transmission distribution is
symmetric -- i.e., $\rho_c(\cT)=\rho_c(1-\cT)$
(Fig.~\ref{fig:Distribution}) -- leading to the saturation of the
third cumulant at high bias.  If the junction is asymmetric, then
the gap at low transmissions opens at $0<\cT<\cT_0$, shifting the
weight of the distribution towards the open channels,
$\rho_c(\cT)<\rho_c(1-\cT)$ for $0<\cT<1/2$, and the high-bias
slope of the third cumulant becomes negative. When one lead is
superconducting, the weight of distribution $\rho_A(R_A)$ for the
symmetric cavity is at low values of the Andreev reflection
probabilities, $\rho_A(R_A)>\rho_A(1-R_A)$ for $0<R_A<1/2$
(Fig.~\ref{fig:Distribution}),
leading to the positive high-bias slope of the third cumulant.
Only for large asymmetries of the junction does the gap that opens
at low $R_A$ prevail and the distribution $\rho_A$ shifts towards
the open Andreev channels and the third cumulant becomes negative
(Fig.~\ref{fig:ThirdCumulant}). It is interesting to note that the
maximally negative slopes of the third cumulant in the normal and
in the Andreev case are approximately equal (if the latter are
corrected for the effective charge). From
Fig.~\ref{fig:Distribution} it is seen that indeed the eigenvalue
distributions are very similar for these values, with the effect
of a superconductor being the change of a junction asymmetry. We
believe it would be interesting to confirm these predictions
experimentally. They provide much more detailed information on the
transmission eigenvalue and Andreev reflection eigenvalue
distributions, which go beyond the information obtained from
conductance and noise measurements.

\section{Conclusion}

We have studied the charge transport statistics in coherent
two-terminal double junctions within the circuit theory of
mesoscopic transport. We have shown that the system of
circuit-theory matrix equations for the Green's function of the
central cavity can be solved, effectively, as a system of scalar
equations independently on the type of the leads.  For an
asymmetric cavity coupled to the leads by quantum point contacts,
the Green's function is expressed in a closed analytical form in
terms of the matrix Green's functions of the leads. The full
counting statistics and the first three cumulants are obtained for
a special case of an open cavity between a superconductor and a
normal metal, at temperatures and bias voltages below the
superconducting gap.

The same results can be obtained by applying the circuit theory
while considering the whole structure as a single connector, with
the cumulant generating function integrated over the distribution
of transmission eigenvalues of the composite junction. This
approach manifestly reveals how the subgap transport in S/N
structures is affected both by the effective charge doubling due
to the Andreev process and by modification of the transmission
properties due to electron-hole correlations introduced by the
superconductor.

For an open cavity, the Fano factor is enhanced with respect to
the corresponding normal-state junction, in agreement with the
recent experimental results by Choi {\it et al.},\cite{art:Choi}
where the high-quality contacts between a cavity and
superconductor have been made. In comparison to the tunnel
coupling, the conductance and Fano factor exhibit opposite trends
as a function of the junction asymmetry, which can be used to
experimentally probe the quality of the contacts. The third
cumulant is strongly affected by the presence of a superconductor.
In contrast to the normal-state case, in which the third cumulant
changes the sign at high enough biases, in the case in which one
lead is superconducting this happens only if the junction is
sufficiently asymmetric. This difference originates from the
skewness\cite{art:Bulashenko} of the Andreev reflection
distribution function, which is in favor of closed Andreev
channels for a moderate asymmetries of the junction.

\section*{Acknowledgments}

We are grateful to Christoph Bruder, Peter Samuelsson, and Oleg
Bulashenko for useful discussions.  This work has been supported
by the Swiss NSF and the NCCR \textit{Nanoscience}.

\appendix
\section*{Appendix}\label{app:A}

Here we show that $\cp_i$ given by Eq.~\eqref{eq:p12} depends only
on the anticommutator of the Green's functions of the leads,
$\cp_i=\cp_i(\aCom{\cG_1}{\cG_2})$, under the assumption that it
is possible to expand $\cG_c$ in a series over the products of
$\cG$ matrices. In this case $\cp_i$ is also a series over $\cG$
matrices, which we denote as $\cp_i=\cp_i (\cG_1,\cG_2)$. In the
following we consider only $\cp_1$, while the reasoning for
$\cp_2$ is analogous. In $\cp_1(\cG_1,\cG_2)$ we separate the term
$\c\varphi(\cG_1\cG_2,\cG_2\cG_1)$, which contains a sum of
products with an even number of $\cG$ matrices, and the term with
the sum of odd-number products. The latter is of the form
$\cG_1\c\phi(\cG_1\cG_2,\cG_2\cG_1)$, where we used
$\cG_1^2=\cG_2^2=1$. As a result,
$
\cp_1(\cG_1,\cG_2)
=
\c\varphi(\cG_1\cG_2,\cG_2\cG_1) + \cG_1
\c\phi(\cG_1\cG_2,\cG_2\cG_1) $. Now we investigate the structure
of $\c\varphi$ and $\c\phi$. First we express $\cG_1\cG_2$ and
$\cG_2\cG_1$ in terms of $\aCom{\cG_1}{\cG_2}$ and
$\com{\cG_1}{\cG_2}$, and then expand
$\c\varphi=\c\varphi(\aCom{\cG_1}{\cG_2},\com{\cG_1}{\cG_2})$ in a
series of $\com{\cG_1}{\cG_2}$. Even powers of
$\com{\cG_1}{\cG_2}$ can be expressed in terms of
$\aCom{\cG_1}{\cG_2}$ by using the identity
$\com{\cG_1}{\cG_2}^2=\aCom{\cG_1}{\cG_2}^2-4$. Thus, $\c\varphi$
is of the form $\c\varphi = \cga + \cgb \com{\cG_1}{\cG_2}$, where
$\cga=\cga(\aCom{\cG_1}{\cG_2})$ and
$\cgb=\cgb(\aCom{\cG_1}{\cG_2})$ depend only on the anticommutator
[and hence they commute with $\cG_1$, $\cG_2$, and
$\cG_c=\cG_c(\cG_1,\cG_2)$]. The same is true for $\c\phi$ --
i.e., $\c\phi = \cga' + \cgb' \com{\cG_1}{\cG_2}$, where
$\cga'=\cga'(\aCom{\cG_1}{\cG_2})$ and
$\cgb'=\cgb'(\aCom{\cG_1}{\cG_2})$. Therefore,
\begin{equation}\label{eq:p1form}
\cp_1 = \cga + \cgb\com{\cG_1}{\cG_2} +
\cga' \cG_1 + \cgb' \cG_1 \com{\cG_1}{\cG_2}.
\end{equation}
On the other hand, $\com{\cG_1}{\cp_1}=0$ in accordance with
Eqs.~\eqref{eq:p12} and \eqref{eq:matrixProp}, and we have
\begin{equation}\label{eq:p1com}
\com{\cG_1}{\cp_1}
= 2\cgb \cG_1 \com{\cG_1}{\cG_2} + 2\cgb' \com{\cG_1}{\cG_2} = 0.
\end{equation}
From Eqs.~\eqref{eq:p1form} and \eqref{eq:p1com} we find
that $\cp_1$ has the following structure:
$
\cp_1 = \cga + \cga' \cG_1.
$
After substituting the expression $\cp_1\cG_1=\cga\cG_1+\cga'$
back into Eq.~\eqref{eq:eqGc} from which $\cG_c$ is to be
obtained, the term $\cga'$ vanishes because it
commutes with $\cG_c$. Thus, when solving for $\cG_c$ we can
assume that $\cp_1=\cga(\aCom{\cG_1}{\cG_2})$ without
loss of generality. A similar consideration holds for $\cp_2$.



\begin{thebibliography}{}
\bibitem{art:BTK}
G.E. Blonder, M. Tinkham, and T.M. Klapwijk, Phys. Rev. B \textbf{25}, 4515 (1982).
%
\bibitem{art:VolkovZaitsev}
A.F. Volkov, A.V. Zaitsev, and T.M. Klapwijk, Physica C \textbf{210}, 21 (1993).
%
\bibitem{art:Khlus}
V.A. Khlus, Zh. Eksp. Teor. Fiz. \textbf{93}, 2179 (1987)
[Sov. Phys. JETP \textbf{66}, 1243 (1987)].
%
\bibitem{art:Takane}
Y. Takane and H. Ebisawa, J. Phys. Soc. Jpn. \textbf{61}, 2858 (1992).
%
\bibitem{art:Beenakker92}
C.W.J. Beenakker, Phys. Rev. B \textbf{46}, R12841 (1992).
%
\bibitem{art:deJongBeenakkerPRB94}
M.J.M. de Jong and C.W.J. Beenakker, Phys. Rev. B \textbf{49},
R16070 (1994).
%
\bibitem{art:ThMartin}
T. Martin, Phys. Lett. A \textbf{220}, 137 (1996).
%
\bibitem{art:Lambert}
C.J. Lambert, J. Phys.: Condens. Matter \textbf{3}, 6579 (1991).
%
\bibitem{art:Datta}
M.P. Anantram and S. Datta, Phys. Rev. B \textbf{53}, 16390 (1996).
%
\bibitem{art:Blatter98}
A.L.~Fauch{\` e}re, G.B.~Lesovik, and G.~Blatter, Phys. Rev. B \textbf{58}, 11177 (1998).
%
\bibitem{art:Imry01}
M. Schechter, Y. Imry, and Y. Levinson, Phys. Rev. B \textbf{64}, 224513 (2001).
%
\bibitem{art:BuettikerPRB01}
K.E.~Nagaev and M.~B{\" u}ttiker, Phys. Rev. B \textbf{63}, R081301 (2001).
%
\bibitem{art:PistolesiPRB01}
F. Pistolesi, G. Bignon, and F.W.J. Hekking, Phys. Rev. B \textbf{69}, 214518 (2004).
\bibitem{art:BuettikerPRB92}
M. B\" uttiker, Phys. Rev. B \textbf{46}, 12485 (1992).
%
\bibitem{art:deJongBeenakker}
M.J.M. de Jong and C.W.J. Beenakker, in \textit{Mesoscopic Electron Transport},
edited by L.P. Kouwenhoven, G. Sch{\" o}n, and  L.L. Sohn (Kluwer, Dordrecht, 1997).
%
\bibitem{art:BeenakkerRevModPhys}
C.W.J. Beenakker, Rev. Mod. Phys. \textbf{69}, 731 (1997).
%
\bibitem{art:BlanterButtikerPhysRep00}
Ya.M. Blanter and M. B\" uttiker, Phys. Rep. \textbf{336}, 1 (2000).
%
\bibitem{nazarov:book}
  \textit{Quantum Noise in Mesoscopic Physics}, edited by
  Yu.V. Nazarov (Kluwer, Dordrecht, 2003).
%
\bibitem{art:SchoelkopfPRL97}
R.J. Schoelkopf, P.J. Burke, A.A. Kozhevnikov, D.E. Prober, and
M.J. Rooks, Phys. Rev. Lett. \textbf{78}, 3370 (1997).
%
\bibitem{art:OberholzerPRB99}
M. Henny, S. Oberholzer, C. Strunk, and C. Sch{\" o}nenberger,
Phys. Rev. B \textbf{59}, 2871 (1999).
%
\bibitem{art:SchoenenbergerPRL01}
S. Oberholzer, E.V. Sukhorukov, C. Strunk, C. Sch\" onenberger,
T. Heinzel, and M. Holland, Phys. Rev. Lett. \textbf{86},
2114 (2001);
%
S. Oberholzer, E.V. Sukhorukov, C. Strunk, and
C. Sch\" onenberger, Phys. Rev. B \textbf{66}, 233304 (2002);
%
S. Oberholzer, E.V. Sukhorukov, and C. Sch\" onenberger,
Nature (London) \textbf{415}, 765 (2002).
%
\bibitem{art:JehlNature00}
X. Jehl, M. Sanquer, R. Calemczuk, and D. Mailly, Nature (London)
\textbf{405}, 50 (2000); X. Jehl and M. Sanquer, Phys. Rev. B
\textbf{63}, 052511 (2001).
%
\bibitem{art:ProberPRL00}
A.A. Kozhevnikov, R.J. Schoelkopf, and D.E. Prober,
Phys. Rev. Lett. \textbf{84}, 3398 (2000).
%
\bibitem{art:HoffmannPRL03}
F. Lefloch, C. Hoffmann, M. Sanquer, and D. Quirion,
Phys. Rev. Lett. \textbf{90}, 067002 (2003).
%
\bibitem{art:Choi} B.-R.~Choi, A.E. Hansen, T. Kontos, C. Hoffmann,
S. Oberholzer, W. Belzig, C. Sch\" onenberger, T. Akazaki,
and H. Takayanagi, Phys. Rev. B \textbf{72}, 024501 (2005).
%
\bibitem{art:ProberPRL03}
B. Reulet, J. Senzier, and D.E. Prober, Phys. Rev. Lett. \textbf{91},
196601 (2003).
%
\bibitem{art:Bomze}
Yu. Bomze, G. Gershon, D. Shovkun, L.S. Levitov, and M. Reznikov,
cond-mat/0504382 (unpublished).
\bibitem{art:NazarovPRL94}
Yu.V.~Nazarov, Phys. Rev. Lett. \textbf{73}, 134 (1994).
%
\bibitem{art:SukhorukovPRL98}
E.V. Sukhorukov and D. Loss, Phys. Rev. Lett. \textbf{80}, 4959 (1998).
%
\bibitem{art:BlanterPRL00}
Ya.M. Blanter and E.V. Sukhorukov, Phys. Rev. Lett. \textbf{84}, 1280 (2000).
\bibitem{art:Andreev}
A.F. Andreev, Zh. Eksp. Teor. Fiz. \textbf{46}, 1823 (1964)
[Sov. Phys. JETP \textbf{19}, 1228 (1964)].
\bibitem{art:BelzigInQNoise}
W. Belzig, in Ref.~\onlinecite{nazarov:book}, p. 463.
\bibitem{art:Muzykantskii}
B.A.~Muzykantskii and D.E.~Khmelnitskii, Phys. Rev. B \textbf{50},
3982 (1994).
%
\bibitem{art:LesovikPRB01}
J. Torr{\` e}s, T. Martin, and G.B. Lesovik, Phys. Rev. B \textbf{63}, 134517 (2001).
%
\bibitem{art:LevitovJETP}
L.S. Levitov and G.B. Lesovik, JETP Lett. \textbf{58}, 230 (1993);
L.S. Levitov, H.W. Lee, and G.B. Lesovik, J. Math. Phys. {\bf 37},
4845 (1996); L.S.~Levitov, in Ref.~\onlinecite{nazarov:book}, p.
373.
%
\bibitem{art:LevitovYakovets}
H. Lee, L.S. Levitov, and A.Yu. Yakovets, Phys. Rev. B
\textbf{51}, 4079 (1995).
%
\bibitem{art:LevitovReznikovCondMat01} L.S. Levitov and M. Reznikov,
 Phys. Rev. B \textbf{70}, 115305 (2004); cond-mat/0111057 (unpublished).
%
\bibitem{art:Bulashenko}
O.M. Bulashenko, J. Stat. Mech. P08013 (2005).
\bibitem{art:Nagaev}
K.E. Nagaev, Phys. Rev. B \textbf{66}, 075334 (2002); K.E. Nagaev,
P. Samuelsson, and S. Pilgram, \textit{ibid.} \textbf{66}, 195318
(2002).
%
\bibitem{art:nazarovANNPHYS99}
Yu.V. Nazarov, Ann. Phys. {\bf 8}, SI-193 (1999).
%
\bibitem{art:BelzigPRL01}
  W. Belzig and Yu.V. Nazarov, Phys. Rev. Lett. {\bf 87}, 197006 (2001).
%
\bibitem{art:Nazarov95}
  Yu.V. Nazarov in \textit{Quantum Dynamics of Submicron Structures},
  edited by H.A. Cerdeira, B. Kramer, and G. Sch{\" o}n
  (Kluwer, Dordrecht, 1995), p. 687.
%
\bibitem{art:NazarovSuperlatt99}
Yu.V. Nazarov, Superlattices Microstruct. \textbf{25}, 1221
(1999).
%
\bibitem{art:NazarovBagretsPRL02}
Yu.V. Nazarov and D.A. Bagrets, Phys. Rev. Lett. \textbf{88}, 196801
(2002).
%
\bibitem{art:BelzigSamuelssonEPL03}
W. Belzig and P. Samuelsson, Europhys. Lett. \textbf{64}, 253 (2003).
%
\bibitem{belzig:01-1}
W. Belzig and Yu.V. Nazarov, Phys. Rev. Lett. {\bf 87}, 067006
(2001).
%
\bibitem{art:belzigPRL04}
P. Samuelsson, W. Belzig, and Yu.V. Nazarov, Phys. Rev. Lett.
\textbf{92}, 196807 (2004).
%
\bibitem{art:SamuelssonPRB03}
P. Samuelsson, Phys. Rev. B \textbf{67}, 054508 (2003).
%
\bibitem{art:HouzetPistolesiPRL04}
M. Houzet and F. Pistolesi, Phys. Rev. Lett. \textbf{92}, 107004 (2004).
%
\bibitem{art:Pistolesi}
G. Bignon, F. Pistolesi, and M. Houzet, cond-mat/0502453
(unpublished).
%
\bibitem{art:SamuelssonButtikerPRL02}
P. Samuelsson and M. B\" uttiker, Phys. Rev. Lett. \textbf{89}, 046601
(2002); Phys. Rev. B \textbf{66}, R201306  (2002).
%
\bibitem{boerlin:02} J. B\"{o}rlin, W. Belzig, and C. Bruder, Phys. Rev.
  Lett. \textbf{88}, 197001 (2002).
%
%



%
\end{thebibliography}
\end{document}